# JGR Space Physics

**Correspondence to:**
S. Haaland,
Stein.Haaland@uib.no

# Characteristics of the Flank Magnetopause: THEMIS Observations

S. Haaland[1,2] , A. Runov[3] , A. Artemyev[3,4] , and V. Angelopoulos[3]

[1]Birkeland Centre for Space Science, University of Bergen, Bergen, Norway, [2]Max-Planck Institute for Solar Systems Research, Göttingen, Germany, [3]Department of Space Physics, University of California, Los Angeles, CA, USA, [4]Space Research Institute, Russian Academy of Science, Moscow, Russia

**Abstract** The terrestrial magnetopause is the boundary that shields the Earth's magnetosphere on one side from the shocked solar wind and its embedded interplanetary magnetic field on the other side. In this paper, we show observations from two of the Time History of Events and Macroscales Interactions during Substorms (THEMIS) satellites, comparing dayside magnetopause crossings with flank crossings near the terminator. Macroscopic properties such as current sheet thickness, motion, and current density are examined for a large number of magnetopause crossings. The results show that the flank magnetopause is typically thicker than the dayside magnetopause and has a lower current density. Consistent with earlier results from Cluster observations, we also find a persistent dawn-dusk asymmetry with a thicker and more dynamic magnetopause at dawn than at dusk.

## 1. Introduction

The terrestrial magnetospause is a current sheet separating the magnetosphere, dominated by the geomagnetic field and hot plasma on one side and the shocked solar wind with its embedded interplanetary magnetic field (IMF) on the other side. The magnetopause plays a key role for the transfer of electromagnetic energy, mass, and momentum from the solar wind into the magnetosphere. It is also host to a number of plasma phenomena and processes, of which magnetic reconnection is perhaps the most intriguing (for a recent update, see, e.g., Burch et al., 2016). The terrestrial magnetopause has therefore been the subject of extensive investigation (see, e.g., reviews in De Keyser et al., 2005; Hasegawa, 2012, and references therein).

Conceptually, the magnetopause can be regarded as a Chapman-Ferraro current sheet (Chapman & Ferraro, 1931; Ferraro, 1952), in which solar wind ions and electrons are deflected in opposite directions by the magnetic gradient in the transition zone between the solar wind and the geomagnetic field. In this theory, the thickness of the current sheet is dictated by the local gyro radii of the ions forming the current and would ideally be 1 gyroradius thick. In reality, this picture is too simple, since there is plasma and magnetic field on both sides of the boundary. Observations have shown that the real magnetopause thickness is typically several ion gyroradii (Berchem & Russell, 1982; Kaufmann & Konradi, 1973; Paschmann et al., 2018).

Due to the focus on processes, in particular those responsible for transfer of momentum and energy across the magnetopause and the associated impact on magnetospheric dynamics, most of the attention has been on the dayside magnetopause near the Sun-Earth line. Although the flanks of the magnetopause have traditionally received less attention, their configuration and dynamics are critical for understanding the transport of magnetosheath plasma to the magnetotail (Wing et al., 2014).

In this paper, we use observations from two of the THEMIS (Time History of Events and Macroscales Interactions during Substorms—see Angelopoulos, 2008) satellites to follow up on results from the Cluster mission presented in Haaland et al. (2014), Haaland et al. (2017), and De Keyser et al. (2017) where observational support and possible explanations for a dawn-dusk asymmetry in magnetopause parameters were presented. In particular, we investigated whether this dawn-dusk asymmetry in magnetopause thickness and velocity is present also in the THEMIS observations. We also discuss the result in context of the more recent survey based on Magnetospheric Multiscale Mission (MMS) magnetopause crossings presented in Paschmann et al. (2018).

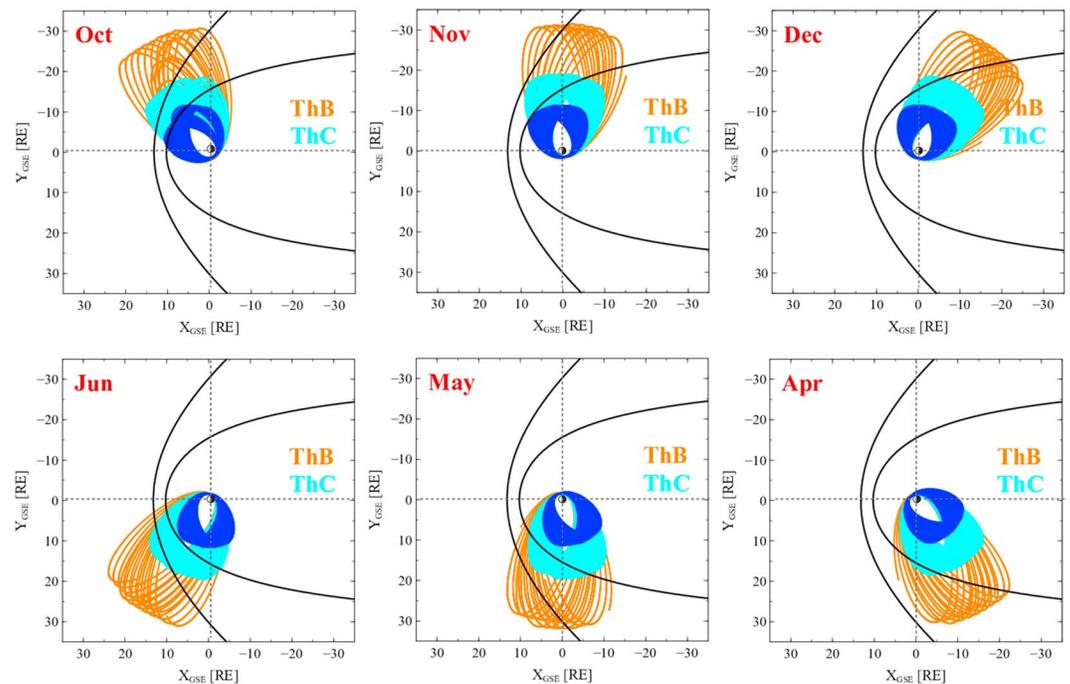

**Figure 1.** THEMIS orbits during October–December (top row) and April–June (bottom row) of the initial (2007–2009) phase of the mission. The dayside is covered during July–September. In this paper, we utilize measurements from THEMIS C (cyan color) and B (yellow color) to characterize the magnetopause, with emphasis on the flanks. Orbits of THEMIS A, D, and E are shown in dark blue color, but they rarely cross the magnetopause during this time interval and are not used in this study.

This paper is organized as follows: In section 2, we present the data basis and provide a brief introduction to the key THEMIS instruments used in this study. Section 3 describes the methodology used to calculate some of the macroscopic parameters of the magnetopause. Section 4 presents the statistical results based on the analyzed data set. Finally, section 5 summarizes the results.

## 2. Observations

THEMIS was launched on 17 February 2007 to study the time evolution of magnetospheric substorms (Angelopoulos et al., 2008). The orbits of the five spacecraft were initially configured so that the inner probes (A, D, and E) covered the region around the flow braking region in the magnetotail just outside geosynchronous orbit and regularly lined up with probe C (apogee around 20 $R_E$) and probe B (apogee around 30 $R_E$) on each side of the near-Earth X-line expected to be formed around 25 $R_E$ downtail during substorms. As a consequence of this configuration and the precession around the Earth, probes B and C will cross the dawn magnetopause flank around October to December, and the dusk flank around April to June as illustrated in Figure 1.

The magnetopause characteristics presented here are based on magnetic field and plasma measurements from THEMIS probes B and C during the period August 2007 to December 2009 and thus encompass three seasons with dawn crossings and two seasons with dusk crossings. Our data set also contains a number of dayside crossings for comparison. Starting in late 2009, the orbits of the spacecraft were lifted and eventually became the ARTEMIS mission (Angelopoulos, 2011), and thus unsuitable for dayside magnetopause studies.

We note that this orbit phasing is almost opposite to that of Cluster, which crossed the dawn flank around May–July and dusk around October–December. Another contrast to the Cluster mission is the orbit; THEMIS has an ecliptic orbit, and the magnetopause crossings take place at low latitudes whereas Cluster has a polar orbit. As pointed out by, for example, Panov et al. (2008), the high-latitude magnetopause can have very different characteristics compared to the low-latitude magnetopause. We also note that the years 2007 to 2009 correspond to solar minimum, whereas MMS phase 1 (years 2015 to 2017) results correspond to high solar activity. THEMIS observations can therefore provide an important complement to the knowledge gained from earlier missions like ISEE and more recent missions like Cluster and MMS.

### 2.1. THEMIS Measurements

To determine macroscopic parameters, and characterize the magnetopause, we use plasma data from the Electrostatic Analyzer (ESA—described in detail in McFadden et al., 2008a) and magnetic field measurements from the Fluxgate Magnetometer (FGM—see Auster et al., 2008). The instrumentation is identical for all THEMIS spacecraft.

The ESA instrument is designed to measure ion and electron distribution functions over the energy range from a few electron volts up to 25 keV for ions and 30 keV for electrons. Full 3-D distributions, from which moments are calculated onboard, utilize the spacecraft spin, and has a resolution of approximately 3 s. In this paper, we primarily use the onboard ion moments (designated MOM in the THEMIS data archives) but consult ion and electron spectra for initial identification of potential magnetopause crossings.

The FGM instrument can measure magnetic fields with up to 64 samples per second, but in this study, we use data with 0.25-s resolution (FGL) to determine magnetopause duration and 3-s spin resolution data (FGS) joined to the ion moments to establish the proper coordinate system and to calculate magnetopause thickness and motion (see section 3.4).

### 2.2. Solar Wind and IMF

The position of the dayside magnetopause is essentially dictated by the balance between the solar wind pressure on one side and the magnetic pressure set up by the geomagnetic field on the other side. At the flanks, the magnetosheath thermal pressure also contributes significantly to the pressure balance. The solar wind velocity and density and thus the dynamic pressure exerted by the solar wind can be highly variable. In addition, local instabilities can excite surface waves that propagate along the flanks (e.g., Kivelson & Chen, 1995). Consequently, the magnetopause at any location is in continuous motion.

In this study we have used time shifted IMF and solar wind parameters (King & Papitashvili, 2005) and geomagnetic disturbance indices downloaded from CDAWEB (Coordinated Data Analysis Web—see https://cdaweb.sci.gsfc.nasa.gov) to monitor these external influences on the magnetopause. Magnetic reconnection is modulated by magnetic shear, that is, the difference between the upstream magnetosheath field and the geomagnetic field. Near noon, the time-shifted IMF from OMNI is probably a good representation of the upstream magnetic field, but at the flanks, there can be significant field line draping (e.g., Sibeck et al., 1990).

## 3. Methodology

The procedure to identify and characterize the magnetopause crossings builds on our experience with Cluster (Haaland et al., 2014) and MMS (Paschmann et al., 2018) and basically consists of the following steps:

1. Identify potential magnetopause crossings by visual inspection of quick-look plots of key parameters. Thereafter, download magnetic field and plasma data for a time period around these potential magnetopause traversals.
2. Establish the current sheet orientation and a local LMN coordinate system.
3. Apply an automated identification of the magnetopause crossing time and duration.
4. Determine magnetopause velocity from a Minimum Faraday Residue (MFR) analysis, and calculate current sheet thickness and current density.
5. Check quality of analysis and store key parameters in database if deemed to be a proper magnetopause crossing.
6. Statistically analyze the records in the database.

These steps are described in some details in the following sections.

### 3.1. Identifying Potential Magnetopause Traversals From Quick-Look Plots

First, we consult the THEMIS mission quick-look WWW pages (at the time of writing, these are available from http://themis.ssl.berkeley.edu/index.shtml). These pages provide low-resolution overview plots of key parameters from the THEMIS mission. We found the plots showing 2-hr overviews of fields, on-board moments, and spectra to be most useful for our purpose. Figure 2 shows an example of a quick-look plot (but reproduced with higher graphics resolution and slightly modified for this paper) with a number of magnetopause crossings.

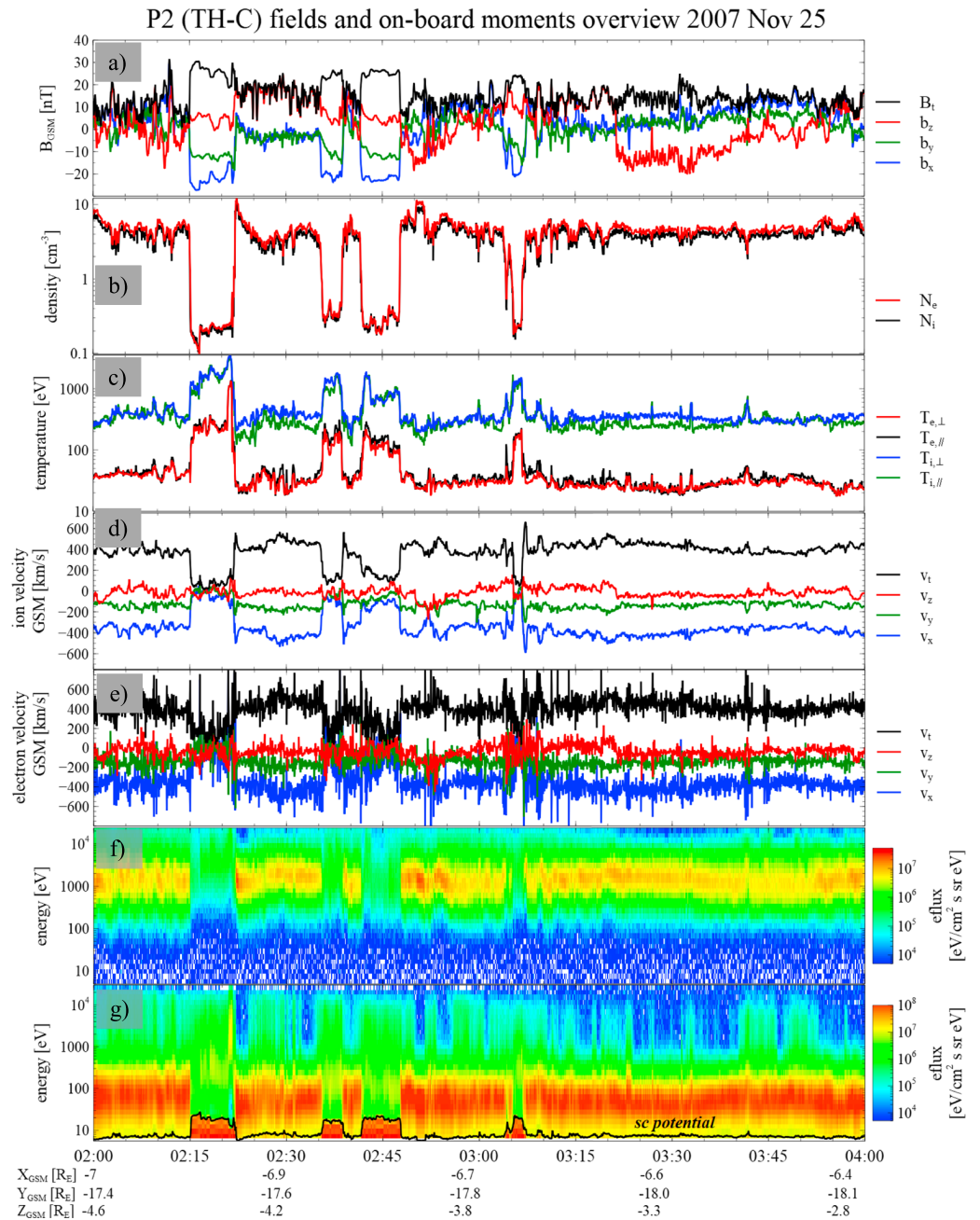

**Figure 2.** Example of quick-look plots used for the initial identification of potential magnetopause crossings. In this 2-hr overview of field and plasma data from THEMIS C on 25 November 2007, a number of magnetopause crossings take place and are manifested as abrupt changes in field and plasma parameters. The panels show (a) magnetic field, (b) plasma density, (c) plasma temperature, (d) ion velocity, (e) electron velocity, (f) electron energy spectrum, and (g) ion energy spectrum.

In the figure, which shows a number of inbound and outbound crossings (i.e., transitions between the magnetosphere and the magnetosheath due to oscillatory motion of the magnetopause) between 02:15 and 03:10 UT on 27 November 2007, the magnetic rotations are easily discernible in panel (a). In the plasma moments, the transitions are characterized by sharp jumps in plasma density (panel b) and temperature (panel c) as the spacraft traverses either from the hot, low density magnetospheric plasma regime to the turbulent magnetosheath, characterized by higher densities and lower temperatures, or vice versa. The flow velocity (ions shown in panel d, electrons in panel e) is typically very low inside the magnetosphere. In the magnetosheath, the flow is often very turbulent near the dayside magnetopause and more laminar with higher flow velocities toward the flanks of the magnetopause.

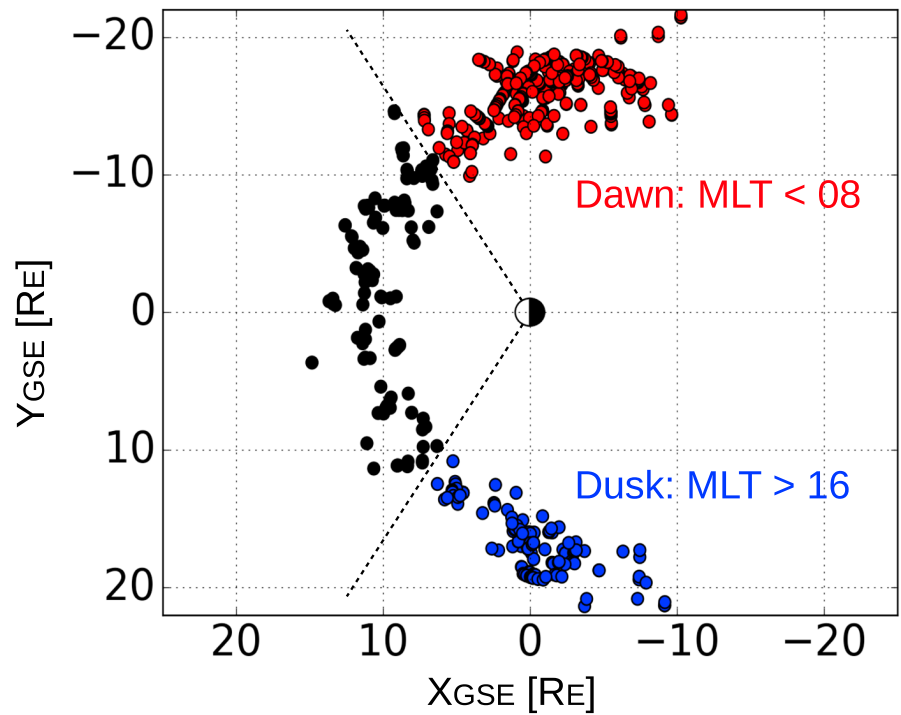

**Figure 3.** Positions of magnetopause crossing positions used in this study. Black dots show dayside crossings, where dayside is defined as crossings within a magnetic local time (MLT) sector between 08 and 16. Dusk crossings (blue symbols) are defined as crossings in MLT sectors 16 to 20, and dawn crossings (red symbols) are defined as positions with MLT 04 to 08. The average radial distance to the dayside crossings is around 11 $R_E$, while the radial distance to both set of flank crossings is around 16 $R_E$.

The transitions from the hot magnetospheric regime to the magnetosheath is also seen in the ion and electron spectra (panels f and g, respectively). The spectra are often useful to distinguish between magnetopause crossings and discontinuities in the magnetosheath or solar wind, which sometimes can have magnetic rotations and jumps in the plasma moments, which can be mistaken for magnetopause crossings. The spectra are also useful to identify crossing times in cases with low magnetic shear, for example, cases with northward IMF.

After having identified potential magnetopause crossings by visual inspection of quick-look plots, we download and plot calibrated magnetic field and plasma data for a 30-min interval around the potential crossing times for each event. In total, data for 2094 time intervals with potential magnetopause crossings in the years 2007–2009 were downloaded. For each event, we also store the solar wind dynamic pressure, the IMF, and geomagnetic indices. After a second visual inspection of plots of the 30-min interval high-resolution data, we discarded some of the events—either because they proved to be magnetosheath or solar wind discontinuities or due to data gaps in some of the measurements. This step reduced the total number of events to 1,297 events, whereof 1,083 crossings from THEMIS C and 214 crossings from THEMIS B.

Figure 3 shows the positions (projected onto the $XY_{GSE}$ plane) of the 1,297 crossings. "Dawn" is defined as crossings taken at magnetic local times (MLT) earlier than 08, and "dusk" is defined as those with MLT after 16. Since our observations have one extra dawn season (autumn of 2007), we have 554 dawn crossings, 398 crossings at dusk, and 365 dayside crossings available for analysis.

Due to its lower apogee and shorter orbital period, there are more magnetopause traversals by THEMIS C than from THEMIS B. The total number of identified THEMIS magnetopause traversals (and eventually the number crossings suitable for the statistical analysis) is significantly lower than the number of crossings in the Cluster results reported by (Haaland et al., 2014; 6,370 crossing during a 10-year period) or the number of MMS crossings in the database constructed by (Paschmann et al., 2018; 2,446 crossings from MMS1 during 2015–2017). When interpreting these numbers, one should keep in mind the short period (2007–2009) and that THEMIS' primary objective was to investigate magnetospheric substorms. In contrast, Cluster was dedicated to study outer magnetospheric boundaries whereas the prime objective of MMS was to study dayside magnetopause reconnection. The lower number of crossings from THEMIS is thus primarily due to the spacecraft orbit and total observation time—not magnetopause identification or detection issues.

### 3.2. Establishing the LMN Coordinate System

From the downloaded magnetic field data, we first construct a LMN coordinate system (Russell & Elphic, 1978) for each event. This coordinate system is constructed by performing a minimum variance analysis (MVAB—see, e.g., Sonnerup & Scheible, 1998) of spin resolution magnetic field measurements over a 5-min interval around the identified crossing. The resulting eigenvectors form a rotation matrix with the three orthogonal unit vectors *L* (orientation of maximum variance), *M* (orientation of intermediate variance), and *N* (orientation of minimum variance). We use this rotation matrix to rotate the high-resolution (four samples per second) magnetic field, which is used to find the crossing time and crossing duration described in the next section.

The *L* axis will typically be well defined since the maximum variance direction is largely governed by the magnetic field inside the magnetosphere. The *N* axis is typically perpendicular to the magnetopause current sheet. Since THEMIS has a near ecliptic orbit, the *N* axis will also typically point in the eclipticic plane, but with opposite $Y_{GSE}$ components at dawn and dusk. We enforce a positive $Z_{GSE}$ orientation of the *L* direction, and an outward pointing *N* direction. The *M* axis completes the right-hand system.

### 3.3. Crossing Time and Duration—Examples

Magnetopause crossing time and duration are determined from the *L* component of the high-resolution magnetic field. As in Paschmann et al. (2018), we smooth the magnetic field measurements using a 3-s boxcar average filter in order to eliminate the effect of small-scale structures inside the main current sheet. Crossing times and durations of the traversals are based on a one-dimensional (1-D) Harris sheet approach (Harris, 1962), in which the current sheet thickness is defined by a 76% change in magnetic field. We define the crossing time as the midpoint (50% level) of the full $B_L$ transition, and the duration as the interval where the magnetic field changes from 12% to 88% of its asymptotic values. The procedure, which is similar to the one we applied on Cluster observations (Haaland et al., 2004, 2014) and later on MMS observations (Paschmann et al., 2018), is illustrated in two examples below.

#### 3.3.1. Example 1—Dayside Crossing on 8 September 2008

Figure 4 shows approximately 2.5 min of measurements from THEMIS C during a dayside magnetopause crossing on 8 September 2008, around 23:40 UT. The magnetic field rotation and changes in plasma density, flow velocity, and temperature are all used to classify and characterize the crossing.

Panel (a) shows the high resolution *L*, *M*, and *N* components and the magnitude of the magnetic field. The crossing time, that is, the time where $B_L$ has reached its 50% level (herafter labeled $t_0$ to be consistent with earlier papers), is indicated by a vertical dashed orange line through panels (a) to (d). Starting at this time, the automated routine then starts searching the boxcar filtered magnetic field forward and backward in time until the times of the 12%, respectively, the 88% level of $B_L$, are reached. These times, $t_{12}$ and $t_{88}$, marked by red dashed lines, define the beginning and end of the current sheet. The time interval between them, $t_{cs} = |t_{88} - t_{12}|$ is the current sheet crossing duration (25 s in this case), which we later use to determine the thickness of the current sheet.

In this example, the magnetic field rotation is fairly well defined, but neither smooth nor monotonous despite our 3-s boxcar filtering. This is typical for almost all of the crossings observed—clean Harris sheet like transitions are rare. The fluctuations in the current sheet indicate either internal spatial structures inside the current sheet or back-and-forth motion of the magnetopause. With a single spacecraft, it is obviously not possible to distinguish between spatial and temporal variations. An accelerating magnetopause implies a deviation from our simple 1-D, stationary model and thus introduce an uncertainty in our determination of velocity and thickness.

We note that the flow velocity (shown in panel c) is enhanced in this case, in particular in the *Y* and *Z* components, as the spacecraft crosses the current sheet. This is most likely a signature of reconnection associated jetting and will be discussed further in section 4.2. We also note that neither the density (panel b) nor the temperature (panel d) immediately reach magnetospheric values after the crossing. This indicates a boundary layer just inside the magnetosphere, possibly formed by the observed reconnection (e.g., Phan et al., 1997; Scholer & Treumann, 1997).

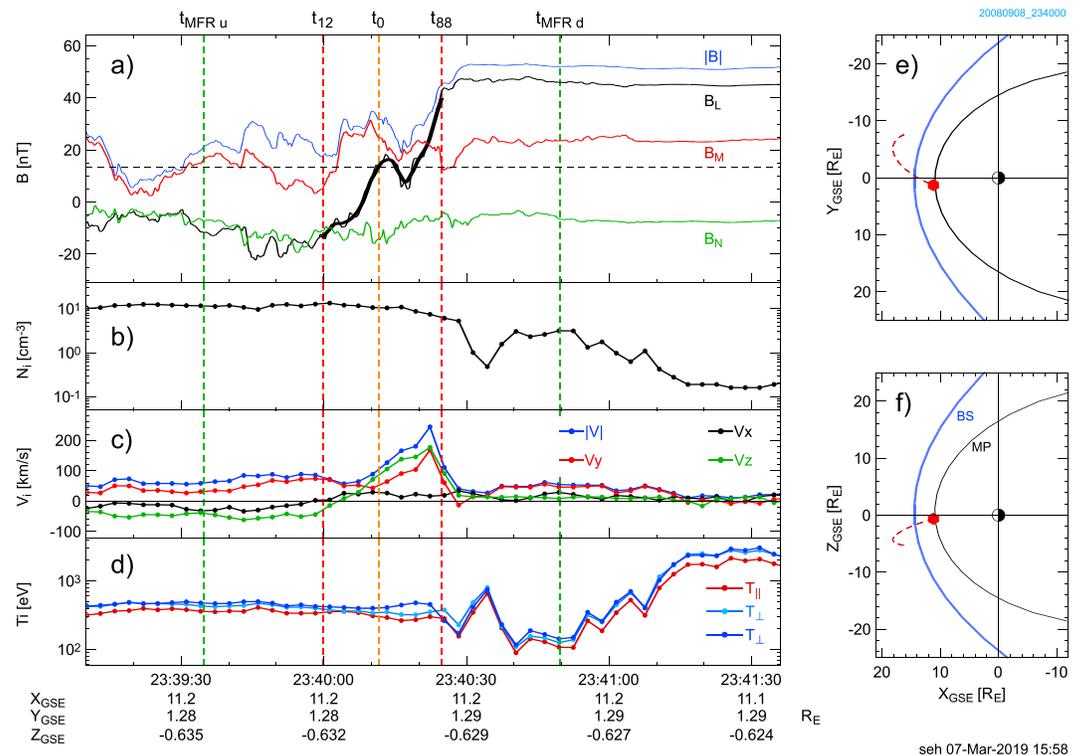

**Figure 4.** Illustration of current sheet crossing times, durations, and analysis intervals for a dayside magnetopause crossing on 8 September 2008 around 23:40 UT. Panel (a), 0.25-s resolution magnetic field in LMN coordinates and the total magnetic field. The thick black curve shows a 3-s boxcar averaged version of the $B_L$ component, which we use to determine duration, and the dashed black horizontal line shows where $B_L$ has reached 50% of its rotation. Panel (b), plasma density. Panel (c), plasma velocity. Panel (d), plasma temperatures. In panels (a) to (d), green vertical lines show the start and stop of time interval ($t_{MFRu}$ to $t_{MFRd}$) used for the HT and MFR analysis, described in section 3.4. The orange vertical line shows the crossing time $t_0$; red vertical lines show the current sheet duration defined as 76% of the total $B_L$ transition. Panel (e), $XY_{GSE}$ projection of spacecraft position (red hexagon) with estimated positions of the magnetopause (MP) and the bow shock (BS) based on Fairfield (1971). Panel (f), as panel (e), but $ZX_{GSE}$ projection.

### 3.3.2. Example 2—Dawnside Crossing on 25 November 2007

Figure 5 shows another example—this time 2.5 min of measurements from an outbound crossing of the dawn magnetopause tailward of the terminator (GSE position [−7 −17 −6] $R_E$) on 25 November 2007. The magnetic field rotation (panel a), along with an increase in density (panel b) and a decrease in temperature (panel d) as the spacecraft leaves the hot and tenuous plasma regime of the magnetosphere and enters the more turbulent magnetosheath region, is readily apparent also in this event.

As with the above dayside event, there are indications of a boundary layer just inside the magnesphere. We note an enhanced density in combined with a reduction in temperatures in the time interval 01:58:06 to 01:58:31 UT. We are not able to directly relate this to any clear reconnection signatures in the form of plasma jetting though, but the bipolar $B_M$ signature may indicate the passage of a magnetic structure.

The flow velocity in the magnetosheath (panel c, right part) exceeds 500 km/s and is primarily in the $-X_{GSE}$ direction as expected for this location. This value is very close to the velocity one would expect from reconnection jets. The local Alfvén velocity, $V_A = B/\sqrt{\mu_0 \rho}$, where B is the observed magnetic field and $\rho$ is the mass density, is also around 500 km/s in this case. High flow velocities are more common at the flanks than at the dayside. It is therefore often difficult to identify reconnection signatures in the form of distinct velocity enhancements. For comparison, the solar wind velocity around this time is just above 600 km/s.

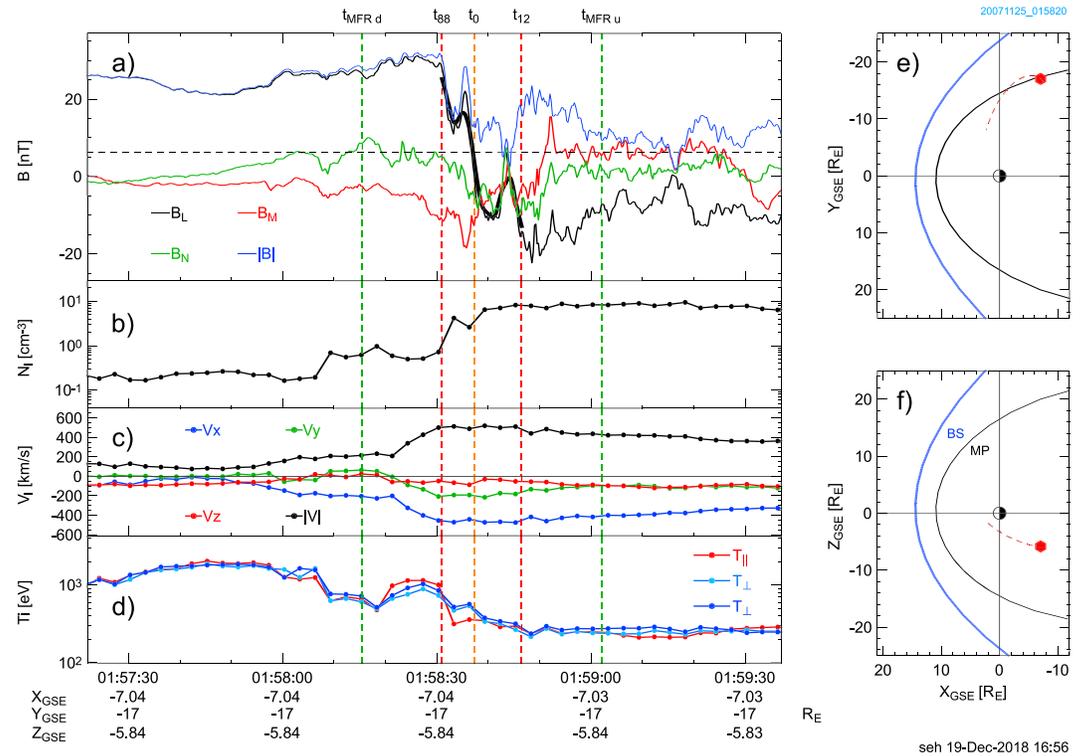

**Figure 5.** Similar to Figure 4, but for a THEMIS C dawn flank crossing on 25 November 2007.

Compared to Figure 4, this crossing has a shorter duration ($t_{cs}$ is ~16 s), but as we will show in the next section, the magnetopause normal velocity is much higher, so the magnetopause is significantly thicker than in the above dayside example.

### 3.4. Magnetopause Velocity and Thickness

To determine the velocity, $V_{MP}$ of the magnetopause current sheet, we apply the MFR (see Khrabrov & Sonnerup, 1998a; Terasawa et al., 1996) method, which returns a frame of reference in which the residual tangential electric field is minimized. MFR also returns a set of eigenvectors that can be used to estimate the orientation (boundary normal, $\vec{n}$) of the magnetopause current sheet as well as a set of eigenvalues that can be used to assess the quality of the determination.

As a check, we also perform a deHoffmann-Teller (HT—see, e.g., Khrabrov & Sonnerup, 1998b; Paschmann & Sonnerup, 2008) analysis, in which the minimization is performed on the electric field, to get the frame velocity. To get the normal velocity from HT analysis, the frame velocity is projected along a suitable boundary normal, in this case the boundary normal, $\vec{n}$, obtained from a constrained minimum variance of the magnetic field (MVAB0).

For both methods, we neglect kinetic effects and assume that the magnetohydrodynamic approximation is valid so that the electric field can be derived from the ion bulk velocity, that is, $\vec{E} = -\vec{V} \times \vec{B}$. Provided that our model of the magnetopause as a stationary magnetohydrodynamic structure is valid, the frame velocity (from either MFR or HT) represents the magnetopause velocity. As in Paschmann et al. (2018) the MFR, HT, and MVAB analyses all use a longer time interval, $t_{MFR} = |t_{MFRu} - t_{MFRd}|$ (where the subscripts $u$ and $d$ refer to upstream and downstream), in this case three times the duration of $t_{cs}$. In Figures 4 and 5 this interval is indicated by vertical green lines. The consistency between the MFR and HT velocities, as well as eigenvalues of the analyses, is used to assess the quality of the velocity calculation.

The magnetopause thickness, $d$, can be calculated from the normal component of velocity and duration: $d = (\vec{V}_{MP} \cdot \vec{n}) * t_{cs}$. With the thickness known, the magnitude of the current density can be estimated from the jump in magnetic field across the current layer: $\mu_0 J = \Delta B/d \simeq \Delta B_L/d$. Using this simplified version of Ampéres law, no detailed information about current direction is possible, though, and one only gets a single value representing the average current density across the entire magnetopause current sheet.

Using the above method, we found a velocity of 17 km/s, a thickness of 430 km (6.5 ion inertial lengths), and a current density of approximately 100 nA/m$^2$ for the dayside example in Figure 4. For the dawn event shown in Figure 5, we found a normal velocity of 180 km/s, a thickness of 2,700 km (34 ion inertial lengths), and a current density of approximately 11 nA/m$^2$.

### 3.5. Quality Check and Error Analysis

Before proceeding to the statistical analysis and characterization, we perform a quality check of the calculations and results. As with any experimental data, there are uncertainties in the measurements, the underlying model as well as statistical spread in the data that need to be considered and handled. We therefore filter the data once more before calculating statistical moments. Table 1 provides details about the definition we have used and filtering we performed on the 1297 events.

1. Measurement uncertainties in the FGM data are probably negligible for this study. Assessments and details are given in Auster et al. (2008). Likewise, we assess that uncertainties in the ESA measurements (discussed in some detail in section 3 of McFadden et al., 2008b) do not strongly influence our results. There will be cases where the energy range of the instrument or spacecraft charging combined with the presence of heavy ions or cold ions skew the plasma moments. The plasma bulk velocity, which is most critical for our study, is less affected than the plasma density or temperature. Density and temperature enter the calculations of the ion inertial length and as a pressure correction in the deHoffmann-Teller calculations (seem e.g., Blagau et al., 2015), respectively.
2. Determination of the LMN coordinate system relies on a simple, near 1-D stationary current sheet as the underlying model and uses MVAB to establish the coordinate system. The $L$ direction and thus the $B_L$ component used to determine the crossing duration are typically well defined. Error estimations for MVAB (see, e.g., equation 8.23 in Sonnerup & Scheible, 1998) are performed as part of our analysis chain and indeed confirm a well-defined $L$ direction in the large majority of cases. However, one should have in mind that these are purely statistical uncertainty estimates.
3. Our definition of thickness as a 76% change in the $B_L$ component is fairly objective, but a potential disadvantage is that this definition does not always work well for cases with low magnetic shear. There will also be cases (in particular on the dayside) where the magnetosheath magnetic field magnitude is larger than the magnetospheric field.
4. As noted in section 3.3, internal structures in the magnetopause current sheet or an accelerating magnetopause also imply a deviation from our simple 1-D, stationary model. Nonconstant motion means less accuracy in the velocity determination and thus a less accurate determination of the magnetopause thickness and velocity.
5. The magnetopause crossing duration and thus the number of samples used for analysis vary from event to event. As in Paschmann et al. (2018), we first determine the duration of the current sheet crossing from the magnetic field profile, $t_{cs}$, then use a longer interval $t_{MFR}$ for the MFR and HT analyses. This ensures plasma samples from both upstream and downstream of the current sheet, but obviously with a variable number of samples. This procedure is obviously open to discussion for a number of cases, but with a large collection of events it is not feasible nor necessarily any better to individually optimize the results for each event.
6. Determination of magnetopause velocity using the MFR technique also assumes a stationary plasma structure. Error estimates for MFR and HT analysis build on the same principle as the above MVAB error analysis. For a detailed assessment of the errors, we refer to the papers by Khrabrov and Sonnerup (1998c), Sonnerup and Scheible (1998), and Sonnerup et al. (2006). In the statistical analysis, we discard all events where MFR fails completely, for example, due to missing data or where the frame determination is deemed to be unreliable, for example, eigenvalue ratio $\lambda_2/\lambda_1 \leq 2$). As in Haaland et al. (2014), events with calculated current sheet thickness less that 150 km, or more than 10,000 km, are also discarded to avoid outliers.
7. Since our analysis chain requires both magnetic field and plasma data, we are only able to include a fraction of the $\approx$6,000 THEMIS magnetopause traversals reported by Plaschke et al. (2009), using high-resolution magnetic field only. By the same token, the time resolution of the THEMIS moments does not allow any detailed study of small-scale structures the like, for example, bifurcated current sheets (Schindler & Hesse, 2008; Runov et al., 2003) or turbulence inside the magnetopause current sheet (Price et al., 2017; Rezeau & Belmont, 2001).

**Table 1**
*Definitions of Dawn, Dusk, and Dayside Regions and Filter Criteria Used to Discard Records Where Reliable Magnetopause Parameters Could Not Be Determined*

| Quality/criteria | Allowed range | Remarks |
| --- | --- | --- |
| Dawn crossings | MLT ≤ 08 | See Figure 3. |
| Dusk crossings | MLT ≥ 16 | |
| Dayside crossings | MLT 08–16 | Paschmann et al. (2018) used $|Y_{GSE}| \leq 10 R_E$. |
| MFR ($\lambda_2/\lambda_1$) | ≥ 2 | Ensure well-determined magnetopause velocity. |
| MP thickness | 150–10,000 km | Remove records with unrealistic magnetopause thickness. |
| Current sheet crossing duration, $t_{CS}$ | ≥6 s | Implies $t_{MFR} \geq 18$ sec; Ensures minimum 5 plasma samples. |
| Shear angle | ≥45° | Identical to Paschmann et al. (2018). |

*Note.* The magnetic shear angle (angular difference between upstream and downstream magnetic field—taken at $t_{MFRu}$ and $t_{MFRd}$, respectively) criteria is only used for the current density calculations to enable comparison with the results from Paschmann et al. (2018).

Important for all of the above potential sources of uncertainties is that the same criteria has been used for all observations, so there should be no bias in the selection. Nor should there be any significant dawn-dusk asymmetries in instrument response. As we will see below, the statistical spread, which is a measure of the true variability in nature, is probably larger than any of the above uncertainties.

## 4. Results

After having performed the above analyses and quality checks for the 1,297 events, we end up with a collection of 576 magnetopause traversals, with key parameters, such as crossing times and duration, magnetopause velocity, current sheet thickness, current density, and a set of measurements at the upstream and downstream of the magnetopause for each event. To characterize the magnetopause and look at specific questions, for example, dawn-dusk asymmetries, we analyze this collection statistically.

### 4.1. Characteristics

Table 2 summarizes the results of the statistical analysis of the collection of events. Here we show averages of distributions of the various key magnetopause parameters for dusk (>16 MLT), dawn (<08 MLT), and the dayside (08-16 MLT) locations. For comparison, we also list some key figures from the recent MMS based study by Paschmann et al. (2018). To check for any significant influence and bias due to external conditions, we calculated corresponding averages of geomagnetic activity indices, IMF values, and the solar wind dynamic pressure. Some of these averages are shown in rows 7–10.

As in Haaland et al. (2014) and Paschmann et al. (2018), we use median as a measure of average when characterizing the distributions. As opposed to mode, the median is unique and robust. Mean values will typically be somewhat higher due to tails in the various distributions, but the dawn-dusk asymmetries are also existent if we use mean to characterize averages. The standard error, $\sigma = s/\sqrt{N}$, where $s$ is the standard deviation and $N$ is the number of crossings, gives an indication of the spread in the data. The ion inertial lengths (a measure of the scale at which ions become demagnetized) given in the table are calculated as $\Lambda_i = d * \sqrt{N_i}/227$, where $N_i$ is the upstream (magnetosheath side) ion density.

Despite differences in instrumentation and epoch (around solar minimum for THEMIS vs. around solar maximum for MMS and over several years for Cluster), the overall results in Table 2 are largely consistent with earlier Cluster and MMS results. THEMIS observations indicate that the dayside magnetopause is thinner than at the flanks, and there is a dawn-dusk asymmetry in thickness and velocity.

If we express the thickness in terms of ion intertial lenghts (row 3), we observe an even larger dawn-dusk asymmetry with the dawn magnetopause flank being almost 80% thicker than the dusk flank. The main reason for this is the much larger average upstream density at dawn ($N_{dawn} \simeq 7$) than at dusk ($N_{dusk} \simeq 4$). On the dayside, we note that THEMIS results suggest a thinner magnetopause than MMS results. One reason for this may be that THEMIS underestimates the density due to the presence of cold ions or limited energy range of the ESA instrument (see section 3.2 of McFadden et al., 2008b).

**Table 2**
*Key Magnetopause Parameters Based on THEMIS B and C Measurements (Columns B–D, Rows 2–7) and Corresponding Solar Wind and Disturbance Parameters (Rows 6–9)*

| | Parameter | B<br>Dawn | C<br>Dusk | D<br>Dayside | E<br>MMS Dayside[a] |
|---|---|---|---|---|---|
| 1 | Number of crossings | 240 | 116 | 182 | 370–2,446[b] |
| 2 | Thickness (km) | 1,407±99 | 1,149±144 | 642±72 | 705 |
| 3 | Thickness ($\Lambda_i$) | 14.9 | 8.0 | 8.3 | 12.6 |
| 4 | Velocity $Vn_{MFR}$ (km/s) | 66.7±4.3 | 49.5±4.5 | 26.2±2.6 | 38.5 |
| 5 | Velocity $Vn_{HT}$ (km/s) | 67.4±4.3 | 50.7±5.1 | 19.2±3.3 | — |
| 6 | Current density (nA/m$^2$) | 16.7±1.8 | 16.3±2.0 | 34.0±4.7 | 543[c] |
| 7 | IMF By (nT) | 0.1±2.1 | 0.8±2.2 | −0.6±3.0 | 1.2[d] |
| 8 | IMF Bz (nT) | −0.3±1.8 | 0.1±1.5 | 0.1±1.7 | −0.3[d] |
| 9 | Solar wind Pdyn (nPa) | 1.5±0.8 | 1.5±0.6 | 1.3±1.0 | 2.9[d] |
| 10 | Dst (nT) | −14±11 | −8±6 | −9±9 | −14[d] |

*Note.* For comparison, dayside values in column E are based on numbers from Paschmann et al. (2018). Values given are medians. Standard errors (see text) are given for THEMIS-based thicknesses, velocities, and current densities, and standard deviations are given for IMF, Pdyn, and Dst.
[a]$|Y_{GSE}| \leq 10 R_E$. [b]Depending on parameter investigated. [c]Peak current derived from high-resolution electron and ion moments. Average currents were much lower. As in Paschmann et al. (2018), we used only included events with shear angle ≥ 45° to calculate the current. [d]Not explicitly listed in Paschmann et al. (2018) but calculated from their database.

From the velocities (rows 4 and 5), we infer a more dynamic dawn flank magnetopause—also consistent with the earlier Cluster results. For individual events, the velocities derived from HT and MFR analyses can differ, but the overall consistency show that the dawn-dusk asymmetry in velocity is real. These results are robust in the sense that adjustments in the filter criteria (Table 1) do not change the overall results.

Current densities derived from THEMIS (row 6, columns B–D) are averages across the magnetopause current sheet, based on difference in magnetic field upstream and downstream, and thus not directly comparable to curlometer-derived results from Cluster or currents based on the high-resolution ion and electron moments from MMS. For both these missions, it was possible to derive current density profiles and give peak current density. In particular, note that Paschmann et al. (2018) only give peak currents based on high-resolution (150-ms cadence) plasma moments—their averages are much lower. In the present THEMIS data set, the average jump in magnetic field is smaller for the dusk flank than for dawn, thus giving almost identical current density estimates despite a thinner dusk current sheet. The higher jump at dawn is primarily due to a higher upstream (magnetosheath) magnetic field at dawn, suggesting dawn-dusk differences in magnetosheath properties rather than intrinsic properties of the magnetopause itself as an explanation.

### 4.2. Reconnection Signatures

The most pronounced observational manifestation of magnetic reconnection in space is plasma jetting—accelerated plasma flows from an active reconnection site. As discussed in section 3.3 and illustrated in Figures 4 and 5, reconnection jetting is easier to identify at the dayside magnetopause than at the flanks. At the dayside, the upstream (magnetosheath) plasma flow is usually more stagnant (lower $|V|$ in our overview plots), so that jetting (with velocities up to the local Alfvén speed) emerges as significant flow velocity enhancements around the crossing time. Toward the flanks the magnetosheath flow has a higher velocity, and accelerated magnetosheath flows with velocities higher than the solar wind can also exist under northward IMF (Erkaev et al., 2011; Lavraud et al., 2007). Reconnection associated jetting at the flanks is therefore difficult to detect by visual inspection.

A more quantitative approach to check for reconnection is to do a Walén test (Walén, 1944) or a Q-test (Sonnerup et al., 2018)—both describing the proportionality between the change in flow velocity and the Alfvén velocity across the magnetopause. A high correlation between these two quantities are expected in rotational discontinuities (RDs) associated with reconnection. Earlier studies, for example, Chou and Hau (2012), Paschmann et al. (2005), and Haaland et al. (2014), have classified events with a well-defined HT frame (HT correlation coefficient ≥ 0.85) and a Walén regression line slope ≥ 0.5 as RDs.

As part of the processing chain, we perform a Walén test of all crossings and record the regression slope and HT correlation coefficient in. Ideally, one would do the Walén only for the outer (i.e., the magnetosheath side of the magnetopause current sheet) part since this is where one would expect to see the RD. However, this would have required individual treatment and optimization of each event. In addition, the limited time resolution of the plasma moments would have restricted such an approach for many of our events since there would have been too few values to correlate in a number of cases. We have therefore done the Walén test using plasma and field data the same time interval, $t_{MFR}$, as that used for the MFR, HT, and MVAB analysis.

In our data set, a well-defined HT frame could be established for 111 crossings at dusk and 276 crossings at dawn. Only 18, respectively, 14 of these crossings, had Walén regression slopes above 0.5. These fractions of cases indicating an RD-like magnetopause are comparable to the results reported by Chou and Hau (2012) and only slightly higher than the Cluster results in Haaland et al. (2014). For comparison, the MMS dayside study by Paschmann et al. (2018) found Walén slopes above 0.5 in around 30% of the cases.

It should be emphasized that we did not perform any optimizations such as tuning the analysis intervals, correct for pressure anisotropy (Blagau et al., 2015), or correct for any effects of composition or cold plasma influence (e.g., Wang et al., 2014). It is also possible that the underlying assumptions (i.e., stationarity, near 1-D structure) for the Walén test are less frequently satisfied at the flanks due to surface waves and turbulent structures inside the magnetopause current sheet. A more concise treatment of each individual event and a higher time resolution in the plasma data would probably have revealed more cases with reconnection signatures than the survey like method used in this study.

### 4.3. Dawn-Dusk Asymmetry

Our investigation of THEMIS magnetopause crossings, summarized in Table 2, reveals a dawn-dusk asymmetry in thickness and motion similar to that reported in Haaland et al. (2014). So why is there a dawn-dusk asymmetry in these key parametres?

#### 4.3.1. External Influences and Asymmetries in Magnetosheath Properties

External influences, like the very different nature of the bow shock at dawn and dusk, result in differences in the upstream magnetosheath properties at dawn and dusk (e.g., Dmitriev et al., 2003). In the Chapmann-Ferraro (CF) picture, a difference in magnetic field and temperature at dawn and dusk will lead to different particle gyro radii at the two flanks, and thus different thicknesses. However, given that the simple CF picture of the magnetopause is an oversimplification (and probably even more so at the flanks), quantitative effects of magnetosheath asymmetries on corresponding asymmetries in magnetopause thickness are difficult to assess.

Walsh et al. (2012) and Dimmock et al. (2016), both using THEMIS data partly overlapping with the time interval used in our investigation, found significant dawn-dusk asymmetries with larger densities on the dawnside than on the duskside. Similar results have also been reported earlier by, for example, Paularena et al. (2001), using observations from the Interplanetary Magnetospheric Explorer 8 (IMP 8) satellite and Němeček et al. (2002) using INTERBALL observations. Our observations are no exception—average upstream plasma moments are different at dawn and dusk also in our data set and largely explains the large difference in the thickness expressed in terms of ion inertial lengths.

None of the above studies were able to identify a single parameter such as the Alfvén Mach number, plasma velocity, or the IMF strength that could exclusively account for the asymmetry observed in magnetosheath properties. The reason for a dawn-dusk asymmetry in density and temperature thus does not seem to have a simple explanation, and Němeček et al. (2002) concluded that the results "bring more questions than answers."

#### 4.3.2. Intrinsic Current Sheet Properties

Recently, De Keyser et al. (2017), inspired by earlier theoretical works by Sestero (1964), used a 1-D kinetic Vlasov-Maxwell model to determine the structure of the magnetopause current layer as function of IMF direction. They noted that the electric field profiles across the magnetopause, that is, the combination of the convection electric ($E_{CONV} = -\vec{V}_{magnetosheath} \times \vec{B}_{magnetosheath}$), and the CF charge separation field ($E_{CF}$) were always different at the two flanks.

The difference in electric field profiles was present regardless of IMF direction, even for cases where the IMF was draped along the flow direction on both sides (and thus $E_{CONV}$=0), but is perhaps best illustrated by considering a southward IMF case: Due to the absence of strong flows inside the magnetosphere, the electric field there will be negligible or zero. On the magnetosheath side, however, there is usually a strong tailward flow, so $E_{CONV}$ is nonzero and will have a component opposite to $E_{CF}$ at one flank, and along $E_{CF}$ at the other flank.

Consequently, magnetosheath ions will be accelerated at dawn and penetrate deeper into the current sheet while electrons will be decelerated and penetrate less deep. Inside the current sheet, there will be a larger separation between ions and electrons and thus a thicker magnetopause current sheet. At dusk, the situation will be opposite; ions and electrons will pulled together and result in a thinner current sheet. The difference of such ion-electron decoupling at dawn and dusk flanks should result in different intensities of Hall (polarization) electric fields (Roth et al., 1996) that affect the magnetopause structure (e.g., Artemyev et al., 2017and references therein).

#### 4.3.3. Magnetopause Surface Waves

Motion of the magnetopause is controlled by a combination of direct solar wind variations, (i.e., mainly changes in the solar wind dynamic pressure) and surface waves like, for example, Kelvin-Helmholtz (KH) waves. At the flanks, the latter probably plays a larger role. One theory (see, e.g., Fadanelli et al., 2018; Kavosi & Raeder, 2015, and references therein) is that waves are exited by local instabilities at the dayside and propagate toward the flanks. Local conditions such as density and velocity shear determine wave parameters such as wavelengths and wave growth rate.

At present, there does not seem to be a clear consensus about whether surface waves are more frequent on the dawn or dusk flank, however (e.g., Dimmock et al., 2017). Simulations (e.g., Nykyri, 2013) indicate that the dawn flank possesses more favorable conditions for KH wave growth. Observations (e.g., Taylor et al., 2012), however, seem to indicate the KH waves are more frequent at the dusk flank.

### 5. Summary

Based on visual inspection of field and plasma parameters from the THEMIS B and C probes, we identified 1,297 potential magnetopause traversals during the period August 2007 until December 2009. Measurements from these intervals were downloaded and analyzed.

We used a Harris sheet approach and defined the magnetopause current sheet crossing duration as the interval where the $B_L$ component of magnetic field rotated from 12% to 88% of its asymptotic values. The frame velocity of the magnetopause was determined from MFR analysis and deHoffmann-Teller analysis of field and plasma data. Reliable crossing durations and magnetopause velocities could be determined for 538 crossings, whereof 240 at dawn, 182 at the dayside, and 116 at dusk.

Although the number of classified flank crossings is far lower than those of Haaland et al. (2014) we observe a similar pattern; the dawn magnetopause is thicker and more dynamic than the dusk flank. The median thickness at dawn was found to be 14,00 km, corresponding to approximately 15 ion inertial lengths. The median velocity at dawn was found to be around 67 km/s. At dusk, the current sheet thickness was found to be 1,150 km, corresponding to eight ion inertial lengths. The median velocity at dusk is around 50 km/s.

We are not able to identify a single, unique mechanism or process responsible for the asymmetry in thickness, but note that a dawn-dusk asymmetry in plasma parameters exists already in the upstream magnetosheath plasma. Nor are we able to identify any single mechanism able to explain the more dynamic nature of the dawn flank magnetopause.

Using the Walén test as a measure, we found that only about 8% of the flank crossings had signatures of reconnection (Walén slopes $\geq$ 0.5). It is unclear whether reconnection is less prevalent than at the dayside or whether this is an issue with the Walén test in this plasma regime.

A more detailed investigation of the underlying reason for the observed dawn-dusk asymmetry would probably require a detailed investigation of the internal structure of the current sheet. The cadence of the THEMIS moments is insufficient for this, but efforts using the MMS mission, which provide plasma moments with resolutions down to 30 ms for electrons and 150 ms for ions, may be able to shed more light on the dawn-dusk asymmetry.

**Acknowledgments**

Research efforts by S. H. were supported by the Norwegian Research Council under Grant 223252; Work of A. R., A. A., and V. A. was supported by NASA Grant NNX16AF84G and NASA contract NAS5-02099. A. A. also acknowledge Russian Foundation for Basic Research, Grant 8-02-00218. We would like to thank the following people specifically: C. W. Carlson and J. P. McFadden for the use of ESA data and K. H. Glassmeier, U. Auster, and W. Baumjohann for the use of FGM data provided under the lead of the Technical University of Braunschweig and with financial support through the German Ministry for Economy and Technology and the German Aerospace Center (DLR) under contract 50 OC 0302. Calculations in this paper have made use of the QSAS science analysis system, provided by Imperial College, London. THEMIS data used in this publication are freely available from http://themis.ssl.berkeley.edu/data/themis/.